\documentclass[preprint]{aastex631}

\shorttitle{Expansion of SNR 0519-69.0}
\shortauthors{Williams et al.}
\graphicspath{{./}{figures/}}

\begin{document}

\title{Evidence for a Dense, Inhomogeneous Circumstellar Medium in the Type Ia SNR 0519-69.0}

\author[0000-0003-2063-381X]{Brian J. Williams}
\affiliation{NASA GSFC, Greenbelt, MD 20771}
\author[0000-0002-9886-0839]{Parviz Ghavamian} 
\affiliation{Department of Physics, Astronomy and Geosciences, Towson, MD, 21252; pghavamian@towson.edu}
\author[0000-0002-5044-2988]{Ivo R. Seitenzahl}
\affiliation{School of Science, University of New South Wales, Australian Defence Force Academy, Canberra, ACT 2600, Australia}
\author{Stephen J. Reynolds}
\author{Kazimierz J. Borkowski}
\affiliation{Department of Physics, North Carolina State University, Raleigh, NC, 27695-0001}
\author{Robert Petre}
\affiliation{NASA GSFC, Greenbelt, MD 20771}



\begin{abstract}

We perform an expansion study of the Balmer dominated outer shock of the SNR 0519$-$69.0 in the LMC by using a combination of new {\it HST} WFC3 imagery obtained in 2020 and archival ACS images from 2010 and 2011.  Thanks to the very long time baseline, our proper motion measurements are of unprecedented accuracy. We find a wide range of shock velocities, with the fastest shocks averaging 5280 km s$^{-1}$ and the slowest grouping of shocks averaging just 1670 km s$^{-1}$. We compare the H$\alpha$ images from {\it HST} with X-ray images from {\it Chandra} and mid-IR images from {\it Spitzer}, finding a clear anti-correlation between the brightness of the remnant in a particular location and the velocity of the blast wave at that location, supporting the idea that the bright knots of X-ray and IR emission result from an interaction with a dense inhomogeneous circumstellar medium. We find no evidence for X-ray emission, thermal or nonthermal, associated with the fastest shocks, as expected if the fastest velocities are the result of the blast wave encountering the lower density ambient medium of the LMC. We derive an age of the remnant of $\leq 670 \pm 70$ yr, consistent with results derived from previous investigations. 

\end{abstract}




\section{Introduction} \label{intro}

Supernova remnants (SNRs) in nearby galaxies with well-known distances, such as the Large Magellanic Cloud (LMC), offer an excellent opportunity for studying the energetics of supernova explosions with great accuracy. Of the young remnants of Type Ia supernovae (SNe Ia), the four Balmer-dominated remnants DEM L71, SNR 0509-67.5, SNR 0519-69.0 and SNR 0548-70.4, have offered some of the best testing grounds for studying such aspects as Type Ia progenitor models \citep{ivo19}, collisionless shock physics \citep{smith91,ghavamian2007a,vanadelsberg08,heng10} and the physics of cosmic ray acceleration \citep[e.g.][]{morlino13}. All four Balmer-dominated SNRs were discovered as X-ray sources by the Einstein Observatory \citep{long1981} and confirmed as SNRs by \cite{tuohy82}, where their Balmer-dominated nature was also established.  The optical emission from Balmer-dominated shocks is produced by collisional excitation of neutral hydrogen atoms as they cross the shock.  There are two populations of neutral atoms, a cold population excited by collisions with post-shock electrons and protons, and a hot population generated by charge exchange between the cold hydrogen atoms and hot protons.  The two components produce broad and narrow H$\alpha$ line emission which can be used to diagnose the shock speed and degree of electron-ion equilibration \citep[e.g.][]{chevalier80C,smith91,ghavamian01,ghavamian2007a,vanadelsberg08,morlino12}.  

In recent years, a small subclass of SNe Ia has been identified as interacting with a circumstellar medium (CSM) in the months to years post-explosion \citep{Silverman13,Graham19}. While the number of members in this class is growing, it is still quite small, and systematic searches in the radio \citep{Chomiuk16,Cendes20} and the UV \citep{Dubay22} have turned up only a few examples. In the SNR field, only Kepler's SNR \citep{reynolds07} in the Galaxy and N103B in the LMC \citep{williams14} have shown evidence of substantial circumstellar interaction (at radii of a few pc and ages of several centuries).

Of all the Balmer-dominated SNRs in the LMC, SNR 0519-69.0 (hereafter, 0519) is one of the less well studied. The light-echo of the supernova that gave rise to 0519 presents as that of a spectroscopically normal SN Ia and the derived explosion age is $(600\pm 200)\,\mathrm{years}$ \citep{rest05}. At least several solar masses of material have been swept-up by the blast wave so far \citep{Borkowski06}. Models taking the combined constraints of the age and the forward and reverse shocks (as traced by the broad [Fe \textsc{xiv}] emission) into account are consistent with a Chandrasekhar-mass progenitor \citep{ivo19}. Searches for surviving companions however have not been able to identify a donor/companion star \citep{edwards12}, but see \citet{Li19} for a proposed candidate. A super-soft X-ray source has been ruled out a progenitor for 0519 by \citet{kuuttila19}, based on the absence of a relic ionization nebular surrounding 0519. \citet{li21} recently reported on the discovery of some higher density knots exhibiting low surface-brightness forbidden line emission in 0519.

Although broad and narrow H$\alpha$ were first detected by \cite{smith91}, only the brightest filaments (corresponding to the slower shocks) were observed in that study.  \cite{ghavamian2007a} detected broad Ly $\beta$ emission in the UV with the Far Ultraviolet Spectroscopic Explorer, from which they estimated a shock velocity of 3600$-$7100 km/s. More recently, \cite{hovey18} used a combination of ground-based optical spectra and space-based HST proper motions of 0519 to estimate the forward shock speed and cosmic ray acceleration efficiency of the SNR. \cite{hovey18} used HST observations separated by 1 year (2010 and 2011) to make their proper motion measurements. In this work, we extend the temporal baseline by an order of magnitude, using new observations obtained in 2020 to substantially reduce the uncertainties on the proper motions of the various filaments.

\section{Hubble Space Telescope Observations} \label{obs}

The data considered here are a combination of existing archival HST images of 0519 acquired in 2010 and 2011 (Program ID GO-12017) with the ACS/F658N instrument/filter combination, along with a new epoch of data acquired in 2020 with WFC3/F657N as part of our approved Chandra program (Program ID GO-15989, which was part of a joint Chandra-HST  program; B. J. Williams, PI).  The  WFC3/UVIS data were obtained on 2020 Jun 21 and Aug 10 with the F657N (H$\alpha$) filter, as shown in Table~\ref{hst_obsn}.  The observations obtained in each exposure were dithered to permit removal of artifacts from the data and (for UVIS) to cover the chip gap, and the appropriate FLASH parameter was set for UVIS to reduce the effects of charge transfer inefficiency.  The combined exposure was 8946 s for the WFC3 data.  The corresponding earlier epoch ACS exposures were 4757 s each for both the 2010 and 2011 observations.  

\begin{deluxetable}{ccccc}
\label{hst_obsn}
\tablewidth{0pt}
\tablecaption{HST/ACS and WFC3 Data Used in Expansion Measurements}
\tablehead{
\colhead {Instrument} &  
\colhead {Dataset} &  
\colhead {Filter} &  
\colhead {Exposure Time (s)} &
\colhead{Observation Date}
}
\startdata
ACS/WFC2     &   JBDQ01010  & F658N  &     3660     &  2010-04-17    \\
ACS/WFC2     &   JBDQ01E5Q  & F658N  &     1097     &  2010-04-17    \\
ACS/WFC2     &   JBDQ02010  & F658N  &     3660     &  2011-04-21    \\
ACS/WFC2     &   JBDQ02Y7Q  & F658N  &     1097     &  2011-04-21    \\
WFC3/UVIS2     & IE6M02020 & F657N  &      3042     &  2020-06-21    \\
WFC3/UVIS2     & IE6M52010 & F657N  &      5914     &  2020-08-10    \\
\enddata
\end{deluxetable}

\begin{figure*}[h]
\center
\includegraphics[width=4in]{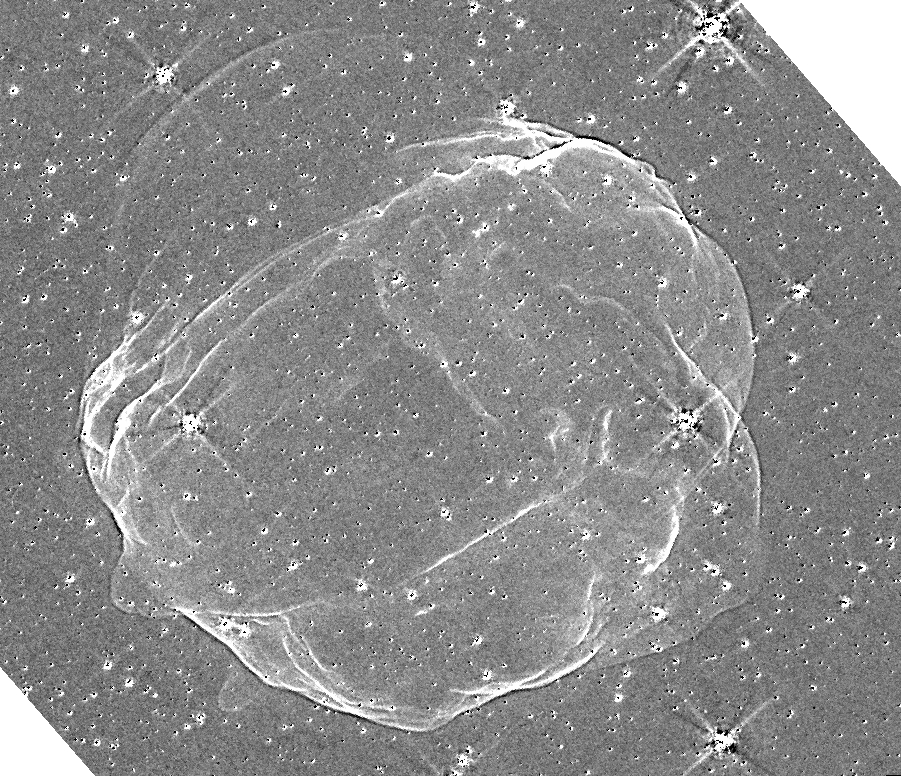}
\caption{Difference image between images of 0519 taken at two epochs: 2010+2011 (ACS F657N images) minus 2020 (WFC3 F658N, rescaled to match ACS).    
The black and white striping along the edge of the blast wave shows sharp, distinct regions of positive and negative pixels, reflecting the substantial expansion movement during the 10-year baseline.}
\label{difference}
\end{figure*}

The HST imagery provides exactly a decade in baseline (2010 to 2020), which, combined with the high shock velocities in 0519 ($\sim$3000 km/s; \cite{hovey18}), should allow a precise, easily measured expansion measurement.  The expansion rate in arcseconds per year, $\omega$, is given by
\begin{equation}
\omega\,=(2.1 \times 10^{-4}) \frac{v_s}{D}
\end{equation}
where $v_s$ is the shock speed (km/s) and $D$ is the SNR distance in kpc.  Using a distance of 50 kpc for the LMC (\cite{clementini03}), shock speeds of $2000 - 3000\,\mathrm{km/s}$ would result in filament displacements of around 0.08\arcsec - 0.1\arcsec over a 10 year baseline, corresponding to 4-5 ACS/WFC3 pixels.   Owing to the small expected displacement ($<$ half a pixel) of the Balmer filaments expected between 2010 and 2011 (as compared to the 2020 WFC3 data), we combined the 2010 and 2011 observations into a single high signal-to-noise image.  Before combining, we utilized the {\it tweakreg} task in Drizzlepac to align the WCS solutions of all WFC3 images and ACS images to one another separately.  The images include numerous well exposed LMC stars, with some near the full-well saturation limit, so the PEAKMAX value  was set to 70,000 electrons during  {\it tweakreg} runs.  A SEARCHRAD parameter was set to 3.0\arcsec to allow a generous offset for search radius.  We then used Astrodrizzle 3.2.1 to combine the 2010 and 2011 ACS images.  As per our calculation above, the Balmer filaments in 0519 will only have moved less than 1/5 ACS pixels between the 2010 and 2011 data, so combining the 2010 and 2011 data will result in far smaller smearing of the filament emission than expected between the ACS data acquisition and the WFC3 data acquisition.  Finally, we used Astrodrizzle to rotate and align the 2020 WFC3 image with the combined 2010+2011 ACS image.  This procedure also resampled the WFC3 data to the ACS pixel scale (0.05$''$ pixel$^{-1}$).  

Examination of the final WFC3 and ACS images showed no residual asymmetries in the stellar PSFs, verifying that the image alignment had been successful.  We checked the absolute astrometric accuracy of the final combined images by measuring centroids for a handful of stars in each field and comparing their coordinates to those of of those stars in the Gaia DR2 catalogue (\cite{Gaia18}).  These coordinates matched to within 0.01$''$-0.02$''$, or less than a quarter of an ACS pixel. In Figure~\ref{difference}, we show a difference image between the two epochs, where virtually all filaments can be easily seen to have moved outward.

\section{Measurements \& Discussion} \label{disc}

Our procedure for measuring the proper motions of the filaments is identical to that used in our previous proper motion studies of various remnants in the optical, radio, and X-ray bands \citep{Williams16,Winkler14,Coffin21}. Briefly, we extract the 1D radial profiles from both epochs (where ``radial" is defined as locally perpendicular to the shock front, which is linear over the small scales that we examine), with uncertainties on each data point, then shift epoch 1 relative to epoch 2, minimizing the value of $\chi^{2}$ when subtracting one profile from the other. Profiles are extracted in pixel space (using the ACS pixel scale size of 0.05$''$), with shifts calculated on a grid of 10000 points with a step size of 0.002 pixels. We smooth the profiles very slightly, using a 1-pixel Gaussian. This has virtually no effect on the profile shapes, but significantly decreases the pixel-to-pixel Poisson noise level. Uncertainties reported are the 90\% confidence limits from the $\chi^{2}$ fitting procedure.

\begin{figure*}
\center
\includegraphics[width=4in]{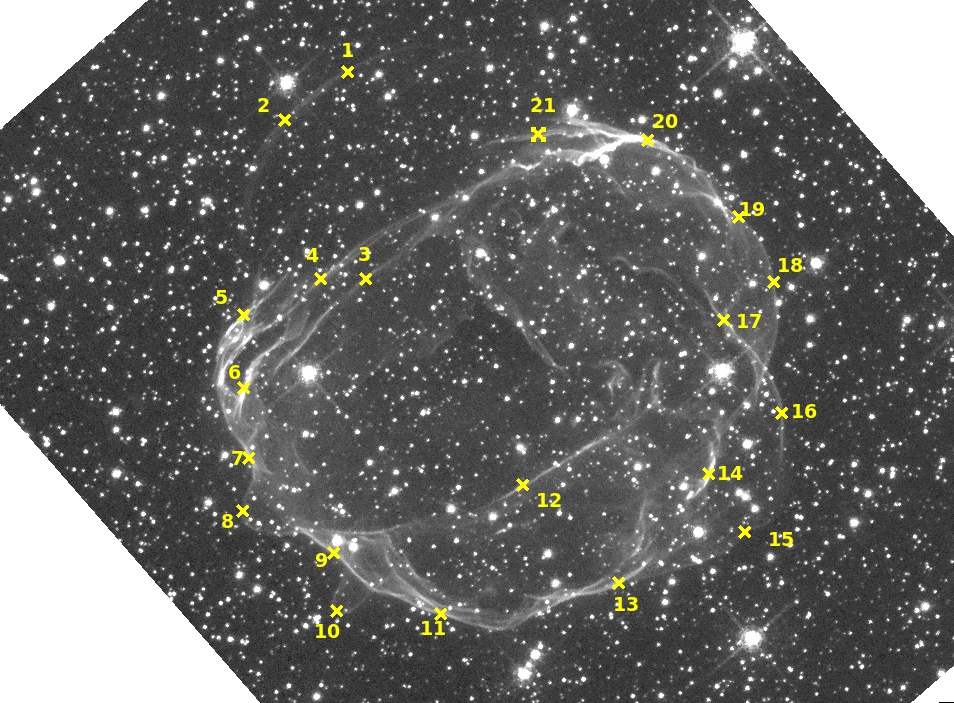}
\caption{The WFC3 F658N image of 0519.  The locations of our expansion measurements around the rim are marked by the numbers.  Proper motion values for the 21 regions are given in Table~\ref{hst_meas}.}
\label{regions}
\end{figure*}

\begin{figure*}
\center
\includegraphics[width=4in]{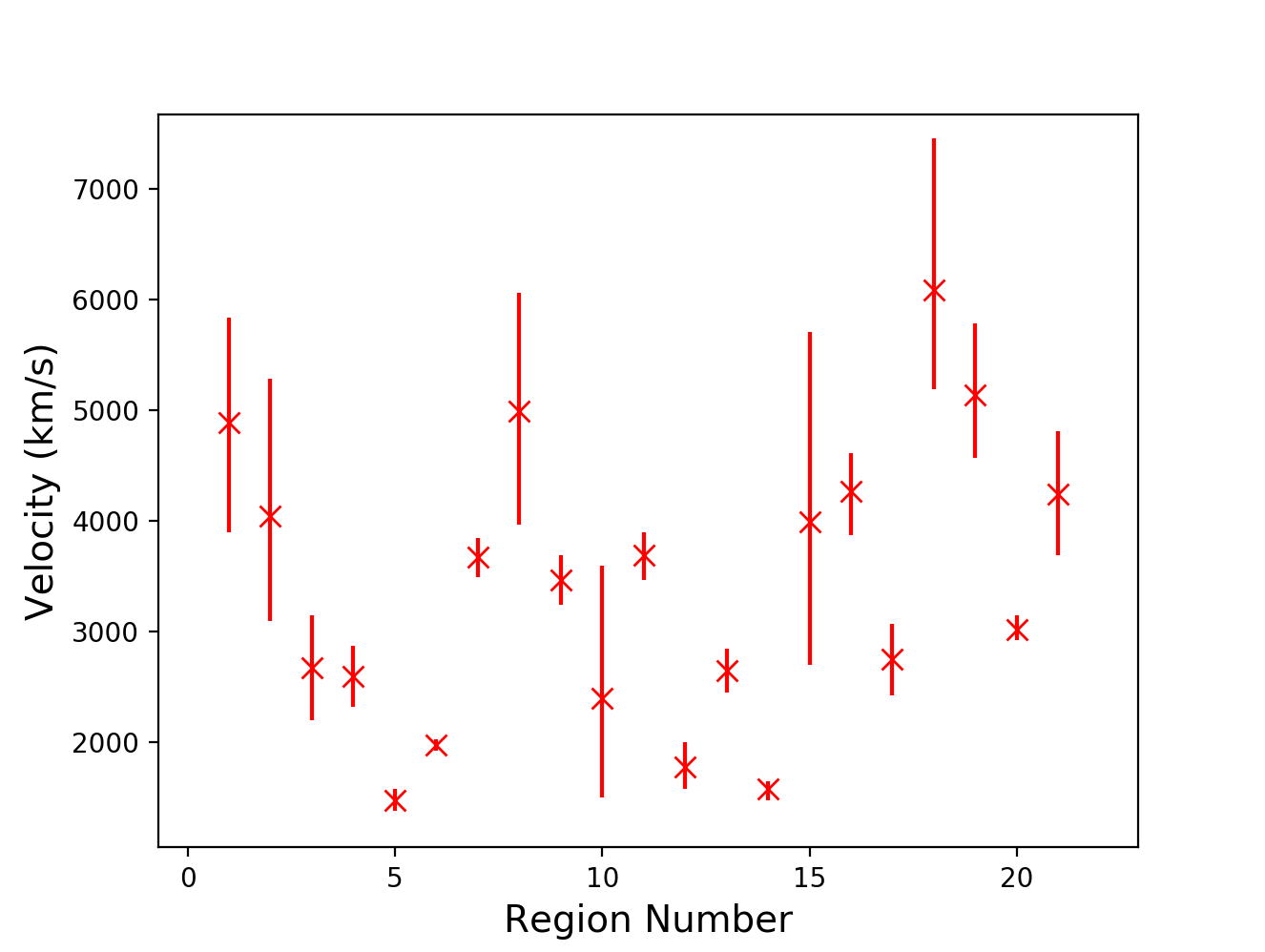}
\caption{Forward shock velocities for 0519 derived from the expansion measurements, shown for the 21 regions around the rim.  Error bars are marked by the vertical lines. In general, the fastest shocks correspond to the faintest filaments, with the slowest shocks correspond to the brightest filaments, consistent with greater deceleration in denser material. See the discussion in the text.}
\label{results}
\end{figure*}

Owing to the high angular resolution of HST and the high signal-to-noise ratio of the filaments, we were able to choose very narrow regions (10 pixels wide, or 0.5$''$) over which to extract our 1D profiles. Such narrow regions ensure that filament curvature is negligible, while also allowing us to easily avoid stars. In Figure~\ref{regions}, we show the 21 regions we selected for analysis. This choice of regions is of course somewhat arbitrary, but we followed two general principles: 1) we chose enough regions around the remnant to get a representative sample size of forward shock velocities; and 2) in places like regions 3$-$6 or 17 and 18, where several distinct filamentary structures are present, we measure each filament's position to search for differences in the expansion velocity. Region 1, 2, 8, and 10 are all located on faint ``blowout'' regions, visible on the image. Results from our measurements are shown in Table~\ref{hst_meas} and Figure~\ref{results}. All values are also converted to velocities, using a distance of 50 kpc to the LMC.

It is apparent from the measurements that there is a spread in velocities present throughout the remnant. In general, the slowest shock speeds (regions 5, 6, and 14) correspond to the brightest filaments, while the faster shock speeds (regions 1, 2, 8, 18, 19) arise from the fainter filaments. Taking the average numbers for those groupings of filaments, we find an average shock velocity for the slower filament group of 1670 km s$^{-1}$, and an average for the faster group of 5280 km s$^{-1}$, corresponding to a factor of around 3 range in shock speed. Under the assumption of pressure equilibrium, where the product of density, $\rho$, and the square of the shock velocity, $v$ (i.e., $\rho v^{2}$) is constant, a factor of 3.1 difference in the shock velocities implies about a factor of 10 difference in the ambient density. 

The flux ratio between the groupings of faint filaments and bright filaments measured above, as measured from the {\it HST} image, is about a factor of 5. Because the H$\alpha$ emission behind the shock scales as the flux of incoming atoms 
$\rho v$, a velocity contrast of 3 and a flux contrast of 5 imply a density contrast of 15. Given the simplicity of these estimates, the rough agreement between the factor of 10 inferred from pressure equilibrium arguments and the factor of 15 inferred from H$\alpha$ brightness is encouraging, and suggests that the remnant is not expanding into uniform density material. 

It is important to note that due to projection effects, the proper motion can only measure the velocity of a filament in the plane of the sky. Getting the true velocity requires having the line-of-sight velocity via spectroscopic measurements. Geometrically, this is likely to be more important for the more internal filaments than those on the rim. In a future work, we will explore spectroscopy of these filaments, which can reveal both the bulk velocity and the widths of the broad spectral components.

\begin{figure*}[h]
\center
\includegraphics[width=2.03in]{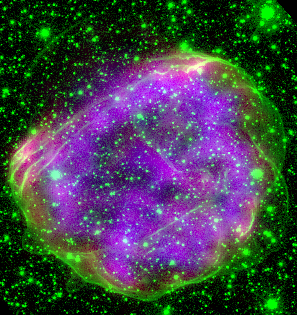}
\includegraphics[width=2.4in]{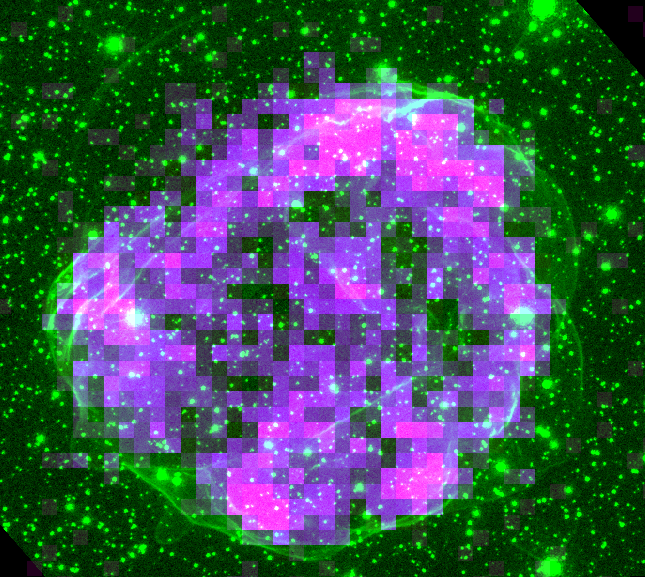}
\includegraphics[width=2.12in]{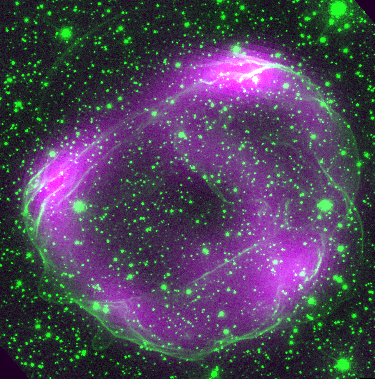}
\caption{{\it Left:} An overlay of the 0.4-0.7 keV X-ray emission from 0519-69.0 in red, the 0.7-0.8 keV emission in blue, and the H$\alpha$ emission in green, from the 2000 observation. The 0.4-0.7 keV energy band is most sensitive to O emission (ambient medium), while the 0.7-0.8 keV band is most sensitive to Fe L-shell emission (primarily ejecta). While both are present throughout, there is a clear enhancement of the O emission in red in the regions corresponding to the brightest, slowest, optical filaments. {\it Middle:} The X-ray emission in hard X-rays from 3.0-6.2 keV is shown in purple, with the H$\alpha$ emission in green. In all energy bands, there is little to no correlation of X-ray emission with the fastest shocks, consistent with those coming from only the lowest density areas. {\it Right:} The de-convolved 24 $\mu$m {\it Spitzer} image is shown in purple, with the H$\alpha$ in green. See discussion in the text.}
\label{optical-xray}
\end{figure*}

One possible explanation for the large density variations on small scales is the presence of a dense shell of circumstellar material which has, in some locations, been penetrated by either the blast wave or clumps of ejecta. Regions 8 and 10 are examples. We have examined in detail the correlation between soft X-ray emission (0.4-0.8 keV) and H$\alpha$ emission. Due to the degradation of {\it Chandra's} soft X-ray response over the years, we use the original {\it Chandra} observation of this remnant from June 2000 (Obs. ID 118, PI S. Holt). In Figure~\ref{optical-xray}, we show images of the remnant in the X-ray and optical bands plotted together. Sharp edges in the X-ray emission match slower shocks with brighter H$\alpha$ emission quite well (e.g., regions 5, 6, 14, and 20). In the left panel of Figure~\ref{optical-xray}, we separate the soft X-ray emission into two components: the 0.4-0.7 keV emission, most sensitive to lines from O, and the 0.7-0.8 keV emission, most sensitive to emission from L-shell emission of Ne-like Fe ions. 

The O emission is most prominent in those regions where the shock speeds are lowest. As was shown in \cite{Borkowski06}, these regions also correspond to the brightest knots of emission seen in {\it Spitzer} 24 $\mu$m images. The authors there concluded that the density in those knots is at least several times higher than in other parts of the remnant. In Figure~\ref{optical-xray}, we show an overlay of the {\it Spitzer} 24 $\mu$m emission with the {\it HST} H$\alpha$ emission. This image has been de-convolved to a resolution of $\sim 2.5$'' (the native resolution of {\it Spitzer's} 24 $\mu$m camera is $\sim 7$'') using the {\it iCore} software package \citep{Masci09}. Similarly to the X-ray emission, the dust emission in IR is brightest where the H$\alpha$ emission is bright and the shocks are slowest. As has been shown in previous studies of this and other remnants with {\it Spitzer} \citep{Borkowski06,williams11}, it is possible to measure the post-shock gas density using emission from warm dust grains, as grain temperature (and thus spectral shape) and overall IR luminosity both scale with density. However, even with de-convolution, {\it Spitzer's} spatial resolution makes this difficult to do for LMC remnants, where the angular sizes of SNRs is small ($\leq$1\arcmin). Future high-resolution observations with {\it JWST} will be critical for gaining a detailed understanding of the density structures encountered by 0519 and other Type Ia SNRs in the LMC.

\begin{figure*}
\center
\includegraphics[width=2.5in]{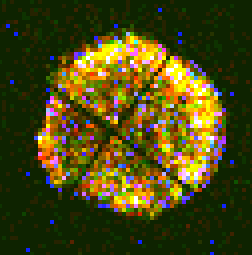}
\caption{Simulated RGB image of Tycho at 50 kpc, as described in text, and based on a 309-second portion of a 2004 Chandra observation (ObsID 3837), not exposure-corrected.  The chip gaps from the ACIS-I instrument are evident.  Red: 0.5 -- 1.8 keV; green, 1.8 -- 4.0 keV; blue, 4.0 -- 6.2 keV.  Note the bluish rim evident around most of the periphery; in the blue energy range, free of strong lines, the synchrotron continuum is expected to be most evident.}
\label{tycho}
\end{figure*}

In contrast, there is little or no X-ray emission present from the faster, fainter filaments (e.g., 1, 2, 8, 15, 18), even in the hard X-rays, as Figure~\ref{optical-xray} shows. If the density contrast really is a factor of 10-15, then the thermal X-ray emission (X-ray emissivity scales as $\rho^{2}$) would be fainter by a factor of 100-200. In fact, this is almost exactly what we measure: when comparing the X-ray count rate in a small area behind region 18 with an identical area behind region 20, we find the X-ray brightness to be higher by a factor of 105 in the latter. It appears that the X-ray emission in 0519 is dominated by interaction with the densest material. We conclude that 0519, in at least some places, is interacting with a dense CSM. This conclusion is supported by the recent work of \cite{li21}, who report that the dense knots seen in forbidden line observations of 0519 are consistent with an origin in a dense CSM. The dense knots seen in that paper correspond to the brightest knots in H$\alpha$ seen in our observations.

All the shock velocities we obtain are well in excess of the $\sim 1,000$ km s$^{-1}$ value often taken as the rough lower limit for shock speeds capable of accelerating electrons to X-ray-synchrotron-emitting
energies $\gtrapprox 1$ TeV. However, X-ray images (shown in Figure~\ref{optical-xray}, which use the recent deep Chandra observations of $\sim 400$ ks) show no evidence of non-thermal filamentary counterparts to the H$\alpha$ filaments that circumscribe 0519. This is perhaps not
surprising; the young Galactic SN Ia remnants Kepler, Tycho, and SN 1006
all show Balmer-dominated shocks in some areas, but typically without
obvious X-ray counterparts. In SN 1006, dominated by nonthermal X-ray
emission, faint H$\alpha$ emission can be followed almost all around the
remnant periphery, and in a few locations, narrow X-ray rims become
evident as H$\alpha$ rim emission weakens, overlapping over tens of
degrees in azimuth.  However, in the NW, where H$\alpha$ emission is strongest, there
is no nonthermal X-ray emission (see images in \cite{Winkler14}). 
The absence of nonthermal X-ray emission at the location of the
H$\alpha$ filaments is somewhat expected on theoretical grounds, in that
the partially neutral upstream medium required for the Balmer emission
is likely to suppress electron acceleration to TeV energies, due to
ion-neutral wave damping \citep{Draine93,Drury96}.  However, a cosmic-ray precursor can be inferred
to be present ahead of Balmer-dominated shocks through detailed optical
spectroscopy (this and related topics are reviewed in \cite{heng10} and \cite{ghavamian13}; also see \cite{knezevic17} for more recent observational results). The
substantial Balmer line emission from 0519, surrounding the entire
remnant, is stronger and more extensive than in Galactic remnants,
perhaps indicating a higher neutral fraction upstream and accounting for
the absence of prominent nonthermal X-ray rims.

One may reasonably ask whether detection of thin rims of nonthermal X-ray emission, such as those seen in Tycho's SNR, is possible at a distance of 50 kpc. To simulate this, we took a single observation of Tycho from 2004 (Obs ID 3837), and assumed its distance to be 2.3 kpc. Translating this to 50 kpc leads to angular scales smaller by a factor of 21.7 and fluxes smaller by a factor of 472. We chose a random piece of the Obs ID 3837 observation (total length 146 ks) that was only 0.309 ks in duration (146/472). We then binned this image by a factor of 22, and generated an RGB image with red (0.5 - 1.8 keV), green (1.8-4.0 keV), and blue (4.0-6.2 keV). We show this image in Figure~\ref{tycho}, where one can see the narrow rims of nonthermal emission in blue. While this is only qualitative, it shows that it should at least be possible to discern nonthermal rims in 0519, if they were present. In a future work, we will examine the X-ray observations in more detail. 

At a distance of 50 kpc, the radius of the remnant of $\sim 15''$ corresponds to a physical radius of $\sim 3.6$ pc. Using the average shock velocities for the fastest filaments reported above (including their uncertainties), we can derive an undecelerated age of $\sim 670 \pm 70$ years. This assumes that the explosion site is in the center of the remnant, an assumption that may not be valid, even for ``circular'' SNRs, though the effect of this is relatively small \citep{Williams13}. Because some amount of deceleration has almost certainly occurred, the real age is somewhat younger than this, which is completely consistent with the age reported in \cite{rest05} of $600 \pm 200$ years. Continued follow-up observations with {\it HST} will further refine estimates of the shock velocity.

\section{Conclusions}

We re-observed 0519-69.0, a young Type Ia SNR in the LMC, using {\it HST} in 2020, a decade after it was first observed in 2010. With this long baseline, we have measured the proper motions of $\sim 20$ filaments in the remnant that constitute a representative sample of the motions present. With the known distance to the LMC, we convert these proper motions to absolute velocities, finding a large range of about a factor of three. Combining our inferred velocities with a morphological study of the remnant at optical, IR, and X-ray wavelengths, we find that the slowest shock velocities result from an interaction of the blast wave with substantially denser material than the average ISM densities in the LMC ($\sim 0.1$ cm$^{-3}$). This dense material likely results from a circumstellar medium. We derive an undecelerated age (an upper limit) for the remnant of 670 $\pm 70$ yr, consistent with previous studies. Young remnants like this should be continuously monitored.

B.J.W. and P.G. acknowledge funding support through HST grant HST-GO-15989.

\begin{deluxetable}{ccccccc}
\tablewidth{0pt}
\tablecaption{Expansion Measurements of 0519 (2010-2020)}
\tablehead{
\colhead {Region Number} &  
\colhead {Proper Motion\tablenotemark{a}} &  
\colhead {PM Lower Limit} &  
\colhead {PM Upper Limt} &  
\colhead {Shock Velocity\tablenotemark{b}} &  
\colhead {SV Lower Limit} &
\colhead {SV Upper Limit} 
}
\startdata
1    & 196   & 156   & 234   & 4890  & 3890  & 5840 \\
2	 & 162	 & 124	 & 212	 & 4040	 & 3090	 & 5290 \\
3	 & 107	 & 88	 & 126	 & 2670	 & 2200	 & 3140 \\
4	 & 104	 & 93	 & 115	 & 2600	 & 2320	 & 2870 \\
5	 & 59	 & 55	 & 63	 & 1470	 & 1370	 & 1570 \\
6	 & 79	 & 77	 & 81	 & 1970	 & 1920	 & 2020 \\
7	 & 145	 & 140	 & 154	 & 3670	 & 3490	 & 3840 \\
8	 & 200	 & 159	 & 243	 & 4990	 & 3970	 & 6060 \\
9	 & 139	 & 130	 & 148	 & 3470	 & 3240	 & 3690 \\
10	 & 96	 & 60	 & 144	 & 2400	 & 1500	 & 3590 \\
11	 & 148	 & 139	 & 156	 & 3690	 & 3470	 & 3890 \\
12	 & 71	 & 63	 & 80	 & 1770	 & 1570	 & 2000 \\
13	 & 106	 & 98	 & 114	 & 2650	 & 2450	 & 2840 \\
14	 & 63	 & 59	 & 66	 & 1570	 & 1470	 & 1650 \\
15	 & 160	 & 108	 & 229	 & 3990	 & 2700	 & 5710 \\
16	 & 171	 & 155	 & 185	 & 4270	 & 3870	 & 4620 \\
17	 & 110	 & 97	 & 123	 & 2750	 & 2420	 & 3070 \\
18	 & 244	 & 208	 & 299	 & 6090	 & 5190	 & 7460 \\
19	 & 206	 & 183	 & 232	 & 5140	 & 4570	 & 5790 \\
20	 & 121	 & 117	 & 126	 & 3020	 & 2920	 & 3140 \\
21	 & 170	 & 148	 & 193	 & 4240	 & 3690	 & 4820 \\
\enddata
\tablenotetext{a}{All proper motions are reported in milliarcseconds over a 10 year baseline.}
\tablenotetext{b}{All shock velocities reported in km/s and assume a distance of 50 kpc for LMC}
\label{hst_meas}
\end{deluxetable}

\bibliography{snr}
\bibliographystyle{aasjournal}



\end{document}